\documentclass[aps,reprint,superscriptaddress,longbibliography,nofootinbib]{revtex4-2}
\pdfoutput=1
\usepackage[utf8]{inputenc}

\usepackage{amsmath,amsthm,amssymb,amsfonts}
\usepackage{scalerel}
\usepackage{mathtools}
\usepackage{graphicx}
\usepackage{stfloats}
\usepackage[normalem]{ulem}
\usepackage[colorlinks = true,
            linkcolor = red,
            urlcolor  = titleblue,
            citecolor = cgreen,
            anchorcolor = blue]{hyperref}
\usepackage{mathrsfs}
\usepackage{bbold}
\usepackage{units}
\allowdisplaybreaks
\usepackage{bm}
\usepackage{braket}
\usepackage{makecell}
\usepackage[nameinlink]{cleveref} 
\usepackage{upgreek}
\usepackage{verbatim}
\usepackage{algorithm}
\usepackage[noend]{algpseudocode}
\usepackage[dvipsnames]{xcolor}
\usepackage{bbm}
\usepackage{array}
\usepackage{multirow}
\usepackage{tabularx}
\usepackage{float}
\usepackage{dcolumn}
\usepackage{slashed}
\usepackage{resizegather}
\usepackage[T1]{fontenc}
\usepackage[export]{adjustbox}

\theoremstyle{definition}

\theoremstyle{plain}

\theoremstyle{plain}
\newtheorem*{prop*}{Property}

\newtheorem*{ctr*}{Conjecture}

\definecolor{titleblue}{RGB}{37,96,167}
\definecolor{cgreen}{RGB}{11,200,10}

\begin{document}

\title{\textcolor{titleblue}{Ergodicity breaking transition beyond qubits}\\
}
\author{Tanay Pathak$^{*,\dagger}$\,\,\href{https://orcid.org/0000-0003-0419-2583}
{\includegraphics[scale=0.05]{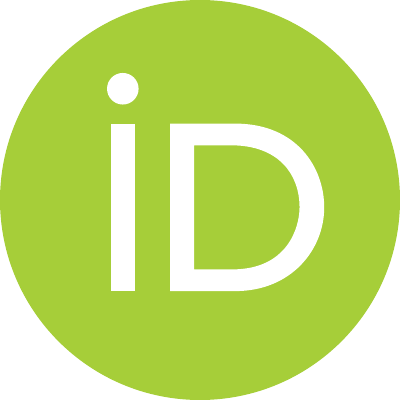}}}
\affiliation{Department of Physics, Kyoto University, Kitashirakawa Oiwakecho, Sakyo-ku, Kyoto 606-8502, Japan}

\author{Abhik Kumar Saha$^{*,\ddagger}$\,\,\href{https://orcid.org/0000-0001-8168-6742}
{\includegraphics[scale=0.05]{orcidid.pdf}}}
\affiliation{Department of Physics, Kyoto University, Kitashirakawa Oiwakecho, Sakyo-ku, Kyoto 606-8502, Japan}

\begin{abstract}
The local Hilbert space dimension is a fundamental parameter in quantum many-body systems, but its role in controlling the ergodicity-breaking transition remains less explored. We systematically study this question by introducing toy models of qudits in zero dimensions that exhibit an ergodicity-breaking transition. Using the hybridization condition, we analytically obtain the dependence of the critical value of the interactions on the local Hilbert space dimension. 
Numerical studies of spectral correlation measures show agreement with the theoretical predictions, hence confirming the existence of the ergodicity-breaking transition in these models. We further introduce a distinct higher spin version of the model and numerically obtain the same critical behavior, demonstrating that the observed critical point is independent of the microscopic structure of interactions considered.
\end{abstract}
\maketitle
\begingroup
\renewcommand{\thefootnote}{*}
\footnotetext{These authors contributed equally to this work.}
\renewcommand{\thefootnote}{$\dagger$}
\footnotetext{Contact author: \href{mailto:pathak.tanay.4s@kyoto-u.ac.jp} {pathak.tanay.4s@kyoto-u.ac.jp}}
\renewcommand{\thefootnote}{$\ddagger$}
\footnotetext{Contact author: \href{mailto:saha.abhikkumar.3k@kyoto-u.ac.jp} {saha.abhikkumar.3k@kyoto-u.ac.jp}}
\endgroup

\emph{Introduction.} Isolated quantum many-body systems possess a fascinating property of thermalization whereby late-time properties of local observables are well described by equilibrium statistical mechanics \cite{PhysRevA.43.2046,PhysRevE.50.888,Deutsch:2018ulr,D_Alessio_2016}. Nevertheless, in the presence of disorder and interactions certain situations can arise when the isolated systems do not thermalize and lead to ergodicity breaking \cite{PhysRevLett.95.206603,Basko_2006,PhysRevB.75.155111,PhysRevB.77.064426,PhysRevB.82.174411,PhysRevB.91.081103,PhysRevB.102.064207}. Studies on ergodicity breaking in interacting quantum systems thus commonly focus on the role of disorder \cite{PhysRevB.82.174411,PhysRevLett.95.206603,Basko_2006,PhysRevB.75.155111,PhysRevB.77.064426,Schreiber_2015,PhysRevB.91.081103,PhysRevLett.111.127201,Smith2016}, spatial dimensionality \cite{Wahl2019,choi_2016}, and the spatial structure of interaction \cite{PhysRevLett.113.243002,PhysRevB.91.094202,PhysRevX.7.041021,PhysRevB.97.214205}. An elementary ingredient, however, is usually kept fixed: the dimension of the local Hilbert space. This quantity determines how rapidly the space of many-body states grows with system size and therefore directly affects the density of states. Its effect should be particularly important in the avalanche mechanism of many-body localization where a finite ergodic region can successively hybridize nearby localized degrees of freedom and thereby grows into a larger thermal region \cite{PhysRevB.95.155129,PhysRevLett.119.150602,PhysRevLett.121.140601,PhysRevResearch.2.033262,PhysRevB.105.174205,PhysRevB.106.L020202,Leonard2023,PhysRevLett.130.250405,PhysRevB.109.L081117,PhysRevB.109.134202}. Such hybridization occurs when the coupling matrix element becomes comparable to or larger than the many-body level spacing of the growing ergodic subsystem. Each successful hybridization enlarges the thermal region and reduces its level spacing, facilitating further growth and potentially leading to a thermal avalanche \cite{PhysRevB.95.155129,PhysRevLett.119.150602,PhysRevLett.121.140601,PhysRevB.105.174205,PhysRevLett.130.250405,Sierant_2025,fvgj-byp7,PhysRevB.110.134204}. Studies concerning the avalanche mechanism, traditionally focus on one-dimensional spin-$1/2$ systems or equally important zero-dimensional toy model such as the \emph{quantum sun model} \cite{PhysRevLett.129.060602,PhysRevB.109.L180201,PhysRevResearch.6.023030,PhysRevB.110.134206,PhysRevB.111.184203,4l42-l7pk}, which provide a controlled setting to analytically determine the ergodicity breaking transition (EBT) in the thermodynamic limit. Understanding the role of local Hilbert space dimensions is thus a particularly important question theoretically as well as experimentally, in view of increasing interest in multilevel quantum systems \cite{PhysRevLett.126.100603,Chen_2025,PhysRevLett.134.113401}, including qudits \cite{Keesling:2018ish,PhysRevA.98.023614,PhysRevLett.122.240402,Mong:2014ova,PhysRevB.105.144306,Pathak:2026oye} and higher-spin degrees of freedom \cite{PhysRevB.103.174203,Hu2024,PhysRevB.106.224205,dabholkar_2024}.

\begin{figure}[H]
    \centering
    \includegraphics[width=\columnwidth]{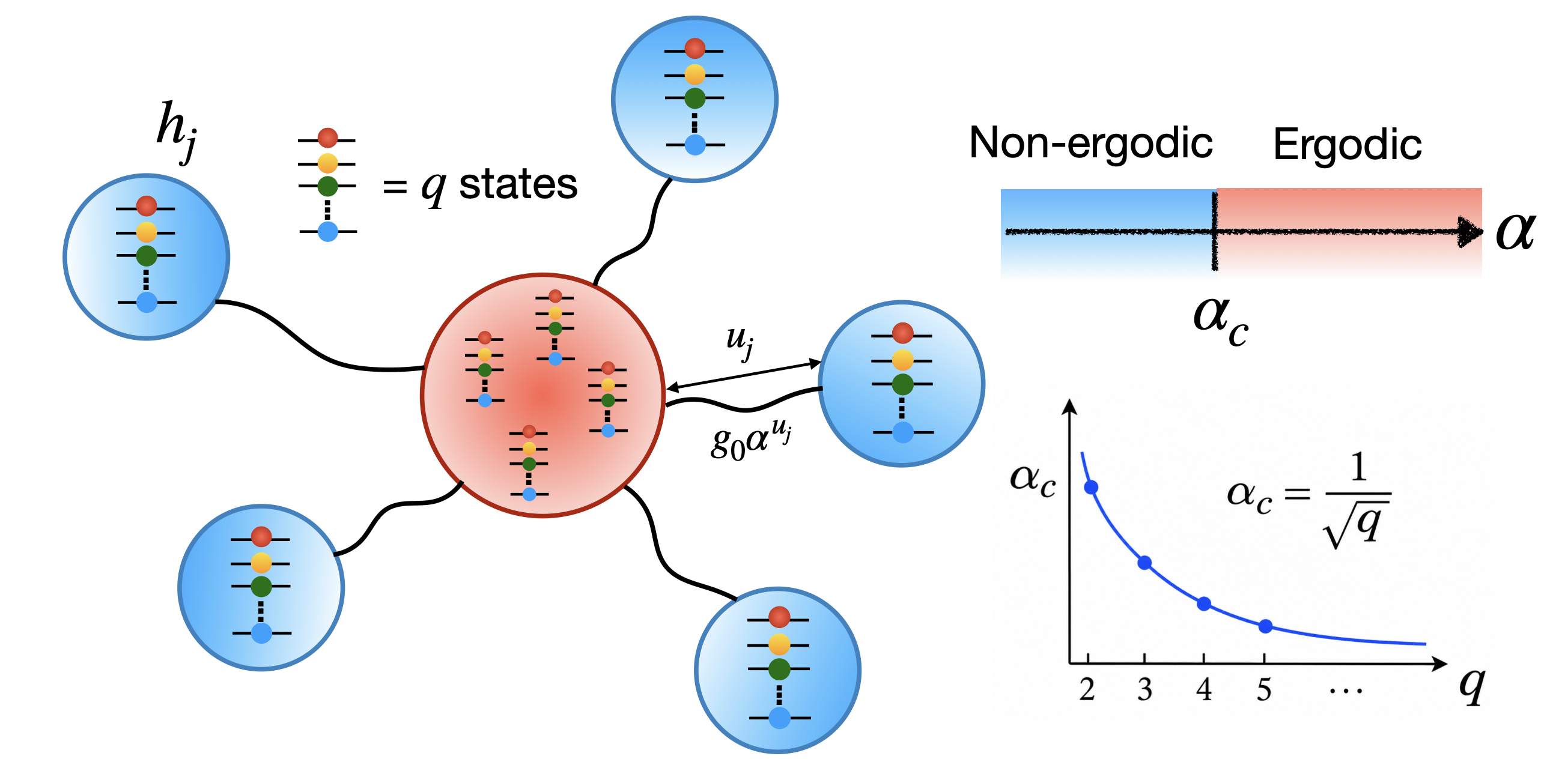}
    \caption{Schematic of the quantum Potts sun model. An ergodic quantum dot (red circle) consisting of $N$ particles and $L$ particles outside the dot, each having $q$ states. Each external particle is subjected to a local field $h_j$ and coupled to a particle in the dot with coupling strength $g_{0} \alpha^{u_{j}}$ that decays with increasing distance $u_j$ from the dot. Increasing $\alpha$ drives the system from non-ergodic to the ergodic regime, with the transition occurring at $\alpha_{c}$. The hybridization criterion predicts a systematic shift of the transition with the local Hilbert-space dimension, $\alpha_{c}=1/\sqrt{q}$.}
    \label{fig:qpsm}
\end{figure}

This raises a fundamental question: \emph{how does the local Hilbert-space dimension control the hybridization process underlying the EBT?} We address this question by introducing a \emph{quantum Potts sun model} (QPSM), a zero-dimensional model of interacting $q$-state degrees of freedom. The model consists of two subsystems: an ergodic quantum dot consisting of $N$ $q$-state particles and $L$ $q$-state particles lying outside the dot, each of which is coupled to a single particle inside the dot (see schematic Fig.~\ref{fig:qpsm}). 
Using a hybridization criterion, we analytically obtain that the critical coupling for the EBT scales with $q$ as $\alpha_{c}=1/\sqrt{q}$. We corroborate this prediction numerically for $q=3$ and $q=4$ state QPSM, using complementary probes of short- and long-range spectral correlations. The average gap ratio captures the crossover from non-ergodic to ergodic regime, while the spectral form factor provides a complementary long-range signature of EBT: it exhibits an approximately universal shape and the ratio of the Heisenberg and Thouless times becomes nearly size-independent at the transition. These independent measures consistently locate the EBT and reveal its systematic shift toward smaller $\alpha$ with increasing $q$, in agreement with the hybridization prediction. We further find that the transition is independent of the underlying microscopic interactions in the model. These results imply that the non-ergodic regime becomes increasingly fragile as the local Hilbert space dimension increases. More broadly, they suggest that the same mechanism may enhance the instability of the MBL phase as the local Hilbert space dimension is increased.


\emph{The model.} Let $\ket{k}, k= 0,1,\cdots, q-1$ denote the basis states. We define the clock operators as: $Z^{m}\ket{l}= \omega^{ml}\ket{l}$ and $X^{m}\ket{l}= \ket{(l-m) \bmod q}$, where $\omega= e^{\frac{2 \pi i}{q}}$ is the $q$-th root of unity, and $Z^{q}= X^{q}= \mathbb{1}$. The system consists of an ergodic quantum dot containing $N$ particles with all-to-all interactions and $L$ particles outside the dot, each having $q$ states. The Hamiltonian is given by \cite{noteham}

\begin{equation}\label{eqn:qsm}
     H = H_{\text{dot}} + \sum_{j=1}^{L} g_0\alpha^{u_{j}}\sum_{a=1}^{q-1}\frac{(X_{n_{j}} X^{\dagger}_{j})^{a}}{\sqrt{q-1}} + \sum_{j=1}^{L} \sum_{a=1}^{q-1}h_{j} \frac{Z^{a}_{j}}{\sqrt{q-1}}.
\end{equation}
The $H_{\rm dot}$ term describes the all-to-all interaction of $N$ particles inside the dot, modelled by a $q^{N}\times q^{N}$ random matrix drawn from the Gaussian orthogonal ensemble (GOE), $H_{\rm dot}=\frac{\gamma}{\sqrt{q^{N}+1}}R$, with  $R=\frac{1}{\sqrt{2}}(M+M^{T})$, where the matrix elements of $M$ are independently sampled from a standard normal distribution. The parameter $\gamma$ sets the energy scale of the dot, while the prefactor $1/\sqrt{q^{N}+1}$ ensures that $H_{\rm dot}$ has unit Hilbert-Schmidt norm $||H_{\text{dot}}||= 1$ at $\gamma=1$, where $||H||^{2}= \mathrm{Tr}(HH^{\dagger})/\mathcal{D}$ and the total Hilbert space dimension is $\mathcal{D}= q^{N+L}$. The second term describes the interaction between the quantum dot and the particles outside the dot. Each external particle $j$ is coupled to a randomly selected particle $n_{j}$ inside the dot. The coupling strength is $g_{0}\alpha^{u_j}$ (we later set $g_{0}=1$), where $\alpha$ controls the decay of the coupling with $u_{j}$ which denotes the distance of the $j$-th external particle from the dot. $u_{j}$ is taken from a uniform distribution, $u_{j}\in [j-\epsilon_j, j+\epsilon_j]$, with $\epsilon_{j}=0.2$. The third term describes the local field acting on the external $q$-state particles. The field strength $h_{j}$ is independently sampled from a uniform distribution, $h_{j} \in [\mathcal{S}-\delta_{\mathcal{S}}, \mathcal{\mathcal{S}} + \delta_{\mathcal{S}}]$ with $\mathcal{S}= 1$ and $\delta_{\mathcal{S}}= 0.5$. The prefactor $1/\sqrt{q-1}$ in the second and the third terms is chosen such that the operator sum $\sum_{a=1}^{q-1}(X_{n_{j}} X^{\dagger}_{j})^{a}$ and $\sum_{a=1}^{q-1}Z^{a}_{j}$  have unit Hilbert-Schmidt norm, ensuring a consistent comparison across different local Hilbert-space dimensions. Apart from energy conservation, the system has no other conservation laws. 

\emph{Hybridization condition.} The hybridization criterion provides a simple estimate of the EBT. We consider the hybridization of the outermost $q$-state particle with the ergodic quantum dot part of the system \cite{PhysRevB.95.155129,PhysRevLett.129.060602}. The outermost particle can hybridize with the ergodic quantum dot when the coupling matrix element connecting it to the many-body states of the dot becomes comparable to or larger than the mean many-body level spacing $\delta E$ of the dot. For the outermost particle, whose distance from the dot is $u_L \simeq L$, the interaction is proportional to $g_0\alpha^{L}\mathcal{O}_{L}$, where $\mathcal{O}_L=\sum_{a=1}^{q-1}
\frac{\left(X_{n_L}X_L^\dagger\right)^a}{\sqrt{q-1}}$ is the normalized Potts coupling operator in the second term. According to the eigenstate thermalization hypothesis (ETH), a typical off-diagonal matrix element of the few-body operator $\mathcal{O}_L$ scales as $\sqrt{\delta E}$, up to a smooth $\mathcal{O}(1)$ structure function \cite{PhysRevA.43.2046,PhysRevE.50.888,Deutsch:2018ulr}. The matrix element responsible for the hybridization therefore scales as $V_L \sim g_0\alpha^L\sqrt{\delta E}$. Hybridization requires $V_L \gtrsim \delta E$. It is therefore convenient to introduce the hybridization parameter
\begin{equation}
    \mathcal{G}_L
    \equiv \frac{V_L}{\delta E}
    \sim \frac{g_0\alpha^L}{\sqrt{\delta E}}.
\end{equation}
To determine its system-size dependence, we note that the relevant ergodic subsystem consists of the quantum dot together with the external particles that have already hybridized with it. Each additional particle increases the Hilbert-space dimension of this subsystem by a factor of $q$. Since the dot size $N$ is kept
fixed, the Hilbert-space dimension therefore grows with $L$ as $\mathcal{D}_L \propto q^L$, and the mean many-body level spacing decreases as $\delta E \propto 1/\mathcal{D}_L\propto q^{-L}$. Substituting this scaling into the hybridization parameter gives $\mathcal{G}_L\sim g_0\left(\alpha\sqrt{q}\right)^L$. Since the prefactor $g_0=1$ is independent of $L$, the hybridization condition
$\lim_{L\to\infty}\mathcal{G}_L \rightarrow \infty$ is satisfied for $\alpha\sqrt{q}>1$, yielding the critical value
\begin{equation}
\alpha_c=\frac{1}{\sqrt{q}}.
\end{equation}
In the following, we establish the existence of the above prediction for finite-size systems using various spectral correlation measures. 


\emph{Spectral correlations.} The first important quantity in our study to see the presence of EBT is the short-range spectral correlations. For this, we consider the average gap ratio of consecutive level spacings \cite{PhysRevB.75.155111}. For $\mathcal{N}$ target ordered eigenvalues in the middle of the energy spectrum, the averaged gap ratio is defined as 
\begin{equation}
 \bar{r}=\frac{1}{\mathcal{N}}\sum_{n=1}^{\mathcal{N}}\frac{\min(\delta_{n},\delta_{n+1})}{\max(\delta_{n},\delta_{n+1})}, \ \delta_{n}=E_{n+1}-E_{n},    
\end{equation}
where $E_n$ denotes the $n$-th many-body energy eigenvalues and
$\delta_n$ is the corresponding level spacing. The resulting
quantity is further averaged over disorder realizations, which we denote
by $\langle r\rangle_{d} \equiv \braket{\bar{r}}_{d}$. For Poisson statistics, $\langle r\rangle_{d} \approx 0.386$, whereas for GOE, $\langle r\rangle_{d} \approx 0.5307$ \cite{PhysRevLett.110.084101}.

\begin{figure}[H]
    \centering
    \includegraphics[width=\columnwidth]{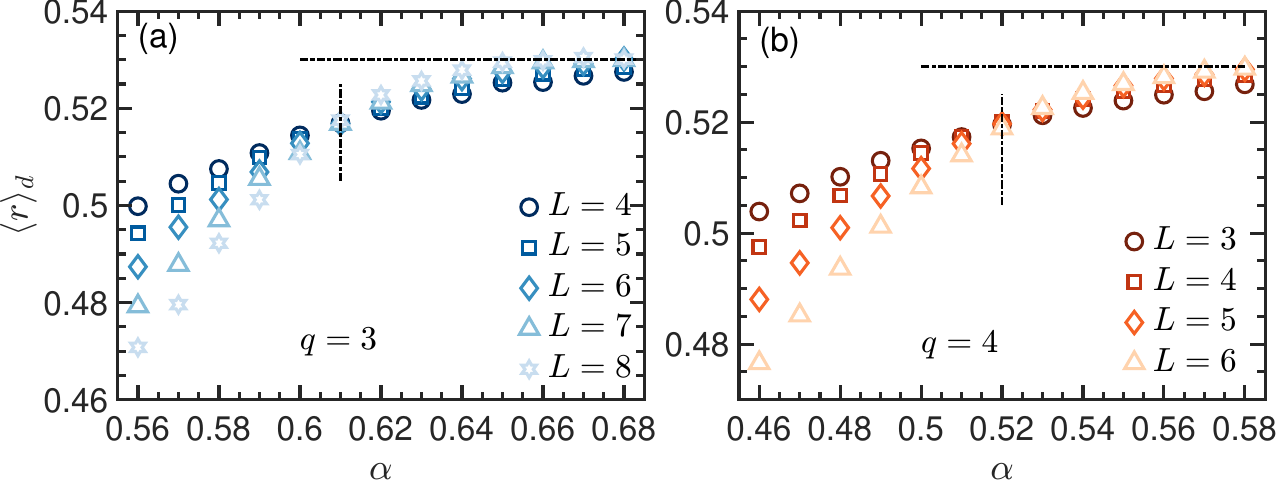}
    \caption{Average gap ratio of consecutive level spacings $\langle r\rangle_{d}$ as a function of $\alpha$ for fixed $N=2$: (a) $q=3$ with $L=4,5,6,7,8$ and (b) $q=4$ with $L=3,4,5,6$. The horizontal dashed line marks the GOE value $\langle r\rangle_{d}\approx0.5307$. The curves become approximately $ L$ independent near $\alpha\approx0.61$ for $q=3$ and $\alpha\approx0.52$ for $q=4$, indicated by the vertical dash-dotted lines. These crossings provide finite-size estimates of the ergodicity-breaking transition.}
    \label{fig:rvalqpsm3_and_4}
\end{figure}

Fig.~\ref{fig:rvalqpsm3_and_4} shows $\langle r \rangle_{d}$ as a function of $\alpha$ for $q=3, 4$ and fix
the quantum-dot size to $N=2$ for both cases. For $q=3$, we consider $L=4,5,6,7$ and $8$ with $5000$, $2000$, $1000$, $500$, and $100$ disorder realizations respectively. For $q=4$, we consider $L=3,4,5$ and $6$, with $10^4$, $5000$, $500$, and $200$ disorder realizations respectively. For our calculations, we consider $\lfloor q^{L+N}/2 \rfloor$ eigenvalues of the middle spectrum. In both cases, $\langle r \rangle_{d}$ increases with $\alpha$, evolving from values closer to the Poisson limit toward the GOE value and indicating a crossover from non-ergodic to ergodic spectral statistics. Moreover, below the transition, $\langle r \rangle_{d}$ decreases with increasing system size, whereas above the transition, it increases toward the GOE value. The curves become approximately size-independent near $\alpha\approx0.61$ for $q=3$ and $\alpha\approx0.52$ for $q=4$, indicating the location of the transition for the finite system sizes considered. These values are close to the hybridization prediction $\alpha_c=1/\sqrt{q}$, which gives $\alpha_c=1/\sqrt{3}\approx0.577$ for $q=3$ and
$\alpha_c=1/\sqrt{4}=0.5$ for $q=4$. A better quantitative estimate of the transition point: $\alpha_{c} \approx 0.610$ and $0.523$ for $q=3$ and $4$ respectively, is obtained later using the data collapse (see Supplemental material \cite{supp}). The finite-size crossings are slightly shifted from the analytical hybridization prediction. The shift could well be assigned to another choice of parameters, such as changing the value of $N$, which is difficult to analyze due to computational limitations in our cases.
\emph{Spectral form factor.} We next examine the spectral correlations on longer energy scales using the spectral form factor (SFF). The SFF is defined as 
\begin{equation}
K(\tau)=\frac{1}{\mathcal{D}}\left\langle\Bigg|\sum_{n=1}^{\mathcal{D}}e^{-2\pi i\epsilon_n\tau}\Bigg|^2\right\rangle_{d},
\end{equation}
where $\{\epsilon_{n}\}_{n=1}^{\mathcal{D}}$ are the unfolded many-body energy levels, $\mathcal{D}=q^{N+L}$, $\tau=t/t_{\rm H}$ is the time measured in units of the Heisenberg time $t_{\rm H}$, and $\langle\cdots\rangle_{d}$ denotes the disorder average. In contrast to the gap ratio, which probes correlations between neighboring energy levels, the SFF probes long-range spectral correlations and contains information about correlations over a broad range of energy scales. It therefore provides a complementary diagnostic of the EBT, and in particular allows us to identify the onset of universal random-matrix behavior through the many-body Thouless time $t_{\rm Th}$. The scaled Thouless time $\tau_{\rm Th}=t_{\rm Th}/t_{\rm H}$ is identified from the time at which the numerical SFF approaches the GOE prediction; $K_{\textrm{GOE}}(\tau)= 2 \tau-\tau \ln(1+2\tau)$ for $\tau\lesssim 1$. Thus, $t_{\rm Th}$ sets the time scale beyond which the system develops universal GOE spectral correlations, whereas $t_{\rm H}$ is set by the inverse mean many-body level spacing. The detailed numerical procedure used to extract $\tau_{\rm Th}$ is given in the Supplementary material \cite{supp}. 
\begin{figure}[H]\centering\includegraphics[width=\columnwidth]{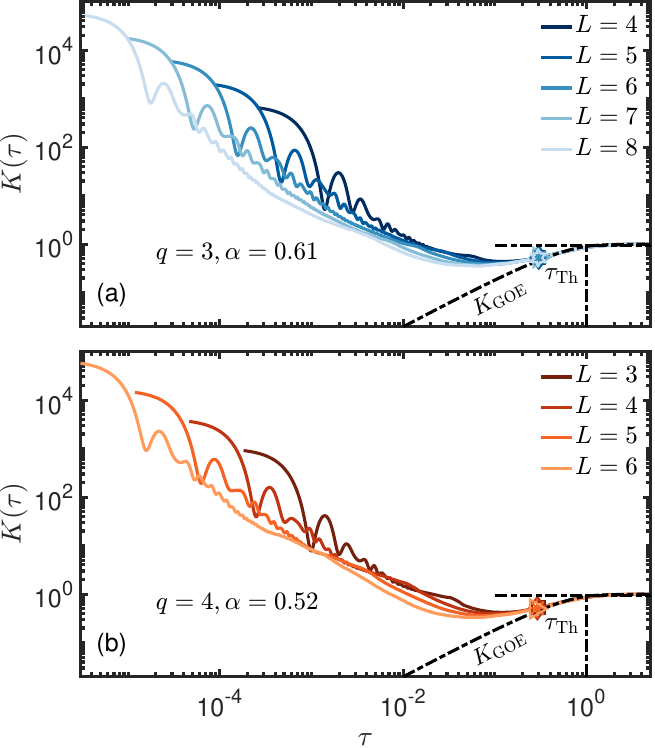}
    \caption{Spectral form factor $K(\tau)$ at the finite-size transition points identified from the gap-ratio analysis for fixed $N=2$: (a) $q=3$ at $\alpha=0.61$ for $L=4,5,6,7,8$, and (b) $q=4$ at $\alpha=0.52$ for $L=3,4,5,6$. The dash-dotted line denotes the GOE prediction $K_{\rm GOE}(\tau)$, while the different markers indicate the extracted scaled Thouless times $\tau_{\rm Th}$ (see Supplementary material \cite{supp}). The vertical dash-dotted line marks the scaled Heisenberg time $\tau_{\rm H}=1$, and the horizontal dash-dotted line indicates the long-time plateau $K(\tau)=1$. Near the transition, the SFF exhibits a weak system-size dependence before approaching the universal GOE form; $K_{\textrm{GOE}}(\tau)= 2 \tau-\tau \ln(1+2\tau)$ for $\tau\lesssim 1$.}\label{fig:sffqpsm3_and_4_critical}
\end{figure}

Fig.~\ref{fig:sffqpsm3_and_4_critical} shows the SFF evaluated at the transition points identified from the gap ratio analysis, $\alpha=0.61$ for $q=3$ and $\alpha=0.52$ for $q=4$, for several system sizes. At short scaled times, the SFF exhibits a clear system-size dependence and nonuniversal structure. With increasing $\tau$, the curves for different system sizes approach one another and move toward the GOE prediction. In the intermediate-time regime, approximately $10^{-2}\lesssim\tau\lesssim10^{-1}$, the size dependence is substantially reduced, while at larger $\tau$, the SFF approaches the universal GOE form before the Heisenberg time $\tau_{\rm H}=1$. 
This weak system size dependence of the SFF at the transition implies that the scaled Thouless time $\tau_{\rm Th}$ is also approximately independent of the system size. The relative scaling of $t_{\rm H}$ and $t_{\rm Th}$ provides a more quantitative criterion for the EBT. In the ergodic regime, $t_{\rm Th}$ grows more slowly with system size than $t_{\rm H}$, so that the ratio $t_{\rm H}/t_{\rm Th}$ increases with system size, implying a long ramp. At the transition, the two time scales acquire approximately the same system-size dependence, implying the absence of a ramp thus leading to 
\begin{equation}
\frac{t_{\rm H}}{t_{\rm Th}}= {\rm const.}
\end{equation}
This motivates considering $\log_{10}(t_{\rm H}/t_{\rm Th})$, 
called the ergodicity breaking parameter (EBP), whose size-independent crossing provides a convenient finite-size criterion for locating the EBT \cite{PhysRevE.102.062144, Suntajs:2021qqr,PhysRevB.107.064205}.
\begin{figure}[h]
    \centering
    \includegraphics[width=\columnwidth]{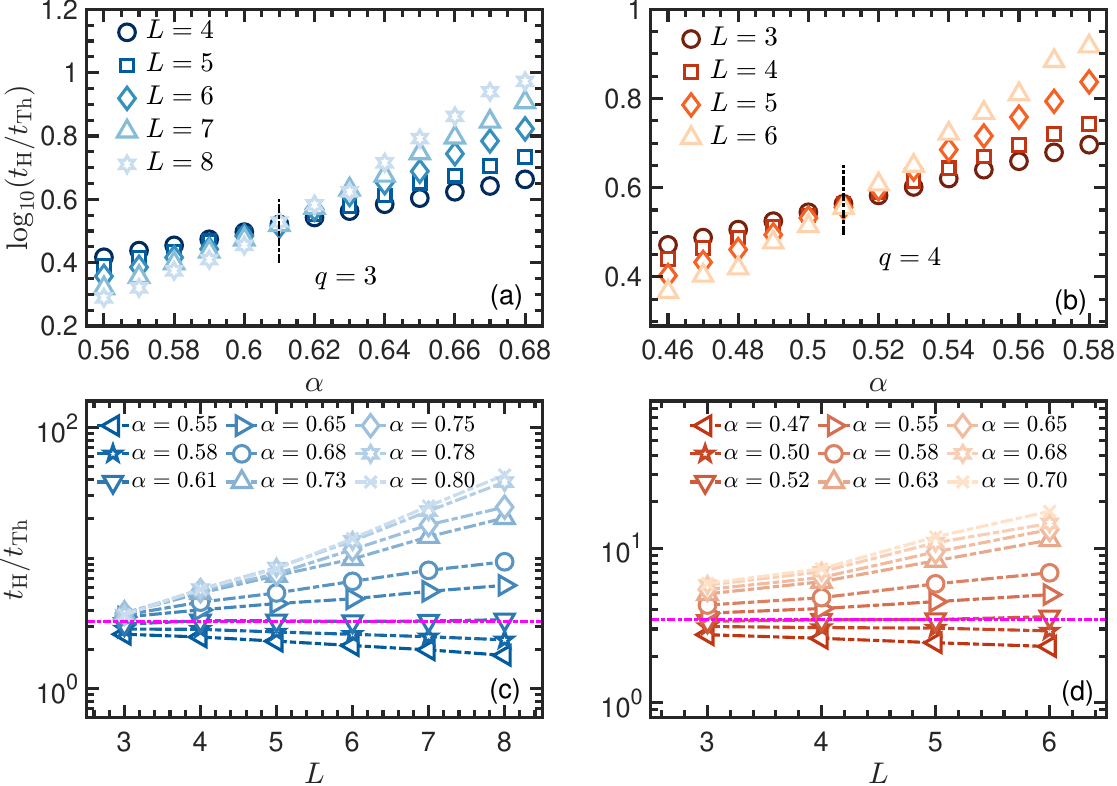}
    \caption{(a,c) Ergodicity-breaking parameter for $q=3$ with fixed $N=2$. (a) $\log_{10}(t_{\rm H}/t_{\rm Th})$ versus $\alpha$ for $L=4,5,6,7,8$, showing an approximately size-independent crossing near $\alpha_{c}\approx0.61$. (c) $t_{\rm H}/t_{\rm Th}$ versus $L$ for different $\alpha$; the ratio decreases below, is nearly constant at, and increases above the transition. (b,d) Corresponding results for $q=4$ with $L=3,4,5,6$, yielding $\alpha_{c}\approx 0.51$. The vertical dash-dotted lines in (a,b) mark the estimated transition points, while the horizontal dash-dotted magenta lines in (c,d) indicate the approximately $L$-independent value of $t_{\rm H}/t_{\rm Th}$ at the transition. The dashed-dotted lines are the fitted exponential curve.}
    \label{fig:sffqpsm3_and_4_ebp}
\end{figure}

Fig.~\ref{fig:sffqpsm3_and_4_ebp} shows $\log_{10}(t_{\rm H}/t_{\rm Th})$ as a function of $\alpha$ for different system sizes. For $q=3$, the curves become approximately size-independent and cross near $\alpha\approx0.61$, whereas for
$q=4$ the corresponding crossing occurs near $\alpha\approx0.51$. These values agree with the transition points identified independently from the gap-ratio analysis. The ratio $t_{\rm H}/t_{\rm Th}$ decreases exponentially with $L$ for $\alpha<\alpha_{c}$, becomes nearly independent of $L$ at $\alpha_{c}$, and increases exponentially with $L$ for $\alpha>\alpha_{c}$. This behavior is consistent with the exponential finite-size dependence found in the quantum sun model \cite{PhysRevLett.129.060602,PhysRevB.109.L180201,PhysRevResearch.6.023030,PhysRevB.110.134206,PhysRevB.111.184203}, with opposite trends on the two sides of the transition. The nearly size-independent ratio $t_{\rm H}/t_{\rm Th}$ at $\alpha_{c}$ therefore provides an additional finite-size signature of the EBT.
The agreement between short-range gap-ratio analysis and the long-range SFF diagnostics is significant because the two probes are sensitive to different spectral scales. The former detects the local level repulsions, whereas the latter probes the development of collective spectral correlations and associated Thouless time. Their consistent transition points therefore show that the observed EBT is not restricted to nearest-neighbor level statistics. This behavior closely parallels the spin-$1/2$ quantum sun model, where a size-independent SFF and a crossing of $t_{\rm H}/t_{\rm Th}$ constitute a key finite-size signatures of the EBT \cite{PhysRevE.102.062144}. The new feature in the present $q$-state QPSM is that same spectral phenomenology persists when the local Hilbert-space dimension is enlarged beyond qubits, while the location of the transition point shifts systematically with $q$. In particular, we find the transition near $\alpha\approx0.61$ for $q=3$ and $\alpha\approx0.51$ for $q=4$, in close correspondence with our results from the gap ratio and the hybridization estimate ($\alpha_c=1/\sqrt{q}$).  A refined estimate using data collapse of EBP gives the transition at: $\alpha_{c} \approx 0.610$ and $0.516$ for $q=3$ and $4$ respectively (see Supplemental material \cite{supp}). 

\emph{Outlook and discussion.} In this work, we demonstrate EBT by introducing a toy model of qudits, dubbed QPSM. The critical value of transition is also consistently identified by three complementary eigenvalue based diagnostics: the size independent crossing of the average gap ratio, the universal shape of the SFF at the transition, and the size-independent crossing of the EBP. For $q=3$ and $q=4$, we find that the transition shifts from $\alpha_{c}\approx 0.61$ to $\alpha_{c}\approx 0.52$ as $q$ is increased. This systematic shift toward smaller $\alpha$ is consistent with the theoretical prediction from the hybridization condition $\alpha_{c}=1/\sqrt{q}$. It is worthwhile to mention that, if we combine the numerical results of the quantum sun model, with $q=2, N=3$ and $\alpha_{c}= 0.75$ \cite{PhysRevLett.129.060602,PhysRevResearch.6.023030}, with our results for $q=3$ and $4$, then the numerical transition values are found to be well described by the fit $\approx 1.05/q^{0.525}$. This numerical fit is in good agreement with the theoretical prediction, thus providing an independent sanity check of our finite size results. We also assert that the scaling of $\alpha_{c}$ is independent of the microscopic form of the interactions. Replacing the clock operators with spin-$1$ and spin-$3/2$ matrix operators (see End matter) yields transition values quantitatively close to those obtained for the corresponding $q=3$ and $q=4$ QPSM, thus confirming this assertion.   

Our results motivate further studies specifically in the context of thermalization and its breaking leading to many-body localization in qudit systems. Results presented in this work imply that ergodic behavior dominates in systems with larger local Hilbert space dimensions and non-ergodic behavior becomes increasingly fragile. More broadly, within the framework of the avalanche mechanism, our results are suggestive of the instability of the many-body localization phase in qudit systems, a prediction that could be well tested both theoretically and experimentally. We also remark that multilevel $q$-state systems have been explored in programmable Rydberg-atom platforms \cite{Keesling_2019}, while Potts-type models have attracted recent interest in the context of topological quantum computation \cite{PhysRevX.5.041040,Alicea:2015hja,PhysRevB.93.125105,trebst2008short}.

\emph{Acknowledgments.---} We are thankful to Masaki Tezuka and Rafa{\l} {\'S}wi{\k{e}}tek for valuable comments. T.~P. gratefully acknowledge support from JST CREST (Grant No. JPMJCR24I2). A.~K.~S. acknowledges support from JSPS KAKENHI Grant Number JP25K00925. The authors acknowledge the computational facilities of the Yukawa Institute for Theoretical Physics, where a majority of numerical computations were performed.
\bibliography{references}

\begin{thebibliography}{70}
\bibitem [{not()}]{noteham}%
  \BibitemOpen
  \href@noop {} {}\bibinfo {note} {We remark that this is not a unique way to construct the Hamiltonian. The current choice in the paper mimics the corresponding Potts model Hamiltonian with $\mathbb{Z}_{q}$ symmetry.}\BibitemShut {Stop}%
\end{thebibliography}
\clearpage
\section*{End Matter}
The hybridization argument discussed in the main text suggests that the location of the EBT is primarily controlled by the growth of the local Hilbert-space dimension rather than by the specific form of the microscopic interactions. To test this directly, we replace the clock operators by conventional spin-$S$ ($S>1/2$) operators. We specifically consider $S=1$ and $S=3/2$, corresponding to local Hilbert-space dimensions $q=2S+1=3$ and $4$, respectively. These models therefore have the same local Hilbert-space dimensions as the $q=3$ and $q=4$ QPSM, but are microscopically distinct: the QPSM is constructed from clock $Z^{q}$ and shift $X^{q}$ operators, whereas the higher-spin models are defined in terms of the higher-spin operators.

The Hamiltonian of the higher-spin quantum sun model is as follows
\begin{equation}
H_{\rm S}=H_{\rm dot}+g_{\rm S}^2\sum_{j=0}^{L-1}g_{0}\alpha^{u_j}S^x_{n_j}S^x_j+g_{\rm S}\sum_{j=0}^{L-1}h_jS^z_j,
\label{eq:spinS_QSM}    
\end{equation}
where, $g_{\rm S}=\sqrt{\frac{3}{S(S+1)}}$ such that $g_{\rm S}^{2}$ and $g_{\rm S}$ ensures unit Hilbert-Schmidt norm of operators $S^x_{n_j}S^x_j$ and $S^z_j$ respectively.

\begin{figure}[bp]
    \centering\includegraphics[width=\columnwidth]{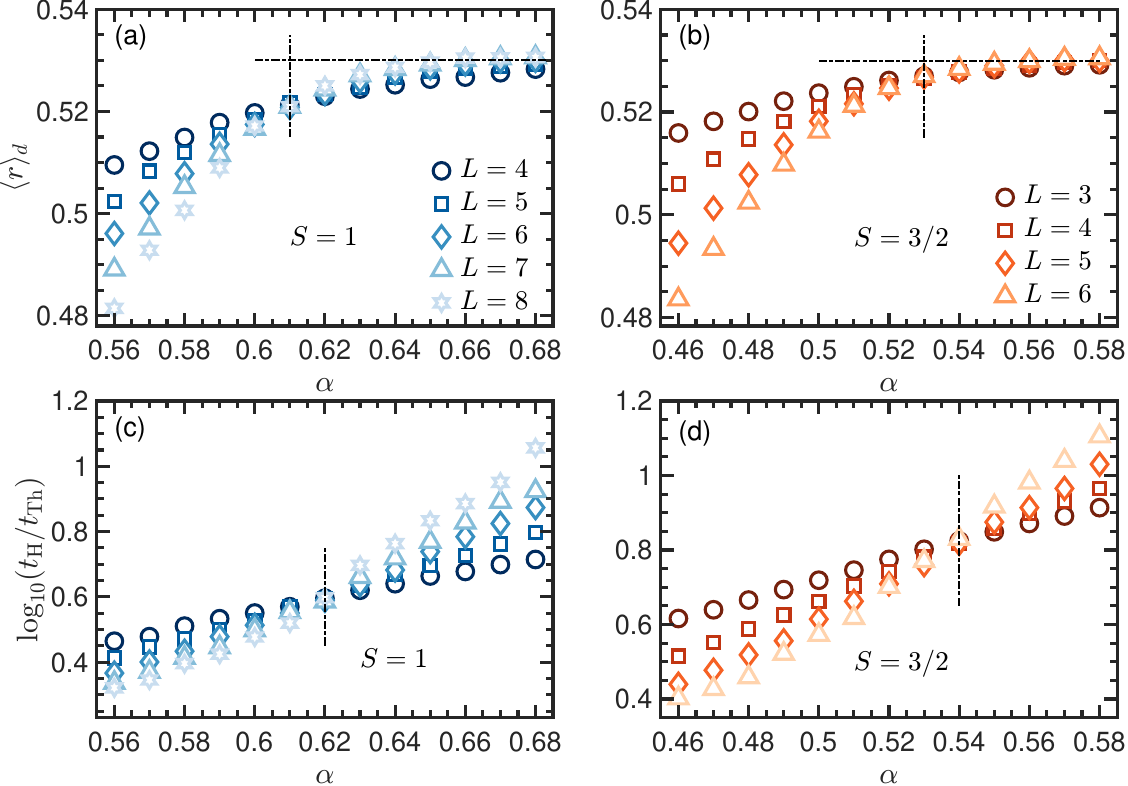}
    \caption{Spectral signatures of the ergodicity-breaking transition for the higher-spin quantum sun models with fixed $N=2$. (a,b) Disorder-averaged gap ratio $\langle r \rangle_{d}$ as a function of $\alpha$ for (a) $S=1$ with $L=4,5,6,7,8$ and (b) $S=3/2$ with $L=3,4,5,6$. The horizontal dashed lines denote the GOE value $\langle r\rangle_{d}\approx0.5307$. (c,d) Corresponding $\log_{10}(t_{\rm H}/t_{\rm Th})$ for (c) $S=1$ and (d) $S=3/2$. The vertical dash-dotted lines mark the estimated transition points.}
    \label{fig:sffratio_high_spin}
\end{figure}

We focus on the same two spectral quantities used to characterize the EBT in the QPSM: the disorder-averaged gap ratio $\langle r\rangle_{d}$ and EBP $\log_{10}(t_{\rm H}/t_{\rm Th})$. Fig.~\ref{fig:sffratio_high_spin} shows both quantities for the $S=1$ and $S=3/2$ cases. For the gap ratio, $\langle r\rangle_{d}$ evolves from Poisson-like to GOE-like statistics as $\alpha$ is increased, while the curves for different system sizes exhibit an approximately size-independent crossing. For $S=1$, the crossing occurs near $\alpha\approx0.61$, close to that found for the $q=3$ QPSM, whereas for $S=3/2$, it occurs near $\alpha\approx0.53$, close to the $q=4$ result.

The same behavior is obtained from the EBP. The curves of $\log_{10}(t_{\rm H}/t_{\rm Th})$ for different system sizes become approximately size independent near $\alpha\approx0.62$ for $S=1$ and $\alpha\approx0.54$ for $S=3/2$. Thus, both the short- and long-range spectral measures locate the transition approximately in the same region and yield transition points close to those of the corresponding QPSM. A more refined finite-size estimate is obtained from the data collapse analysis (see Supplemental material \cite{supp}). From the gap-ratio data collapse, we obtain $\alpha_{c}=0.610$ for $S=1$ and $\alpha_{c}=0.528$ for $S=3/2$. The independent collapse of $\log_{10}(t_{\rm H}/t_{\rm Th})$ gives $\alpha_c=0.618$ and $\alpha_c=0.537$, respectively. These values are close to one another and, importantly, remain close to the corresponding QPSM estimates. 

These results provide a direct test of the hybridization picture. Despite the different forms of operators in the QPSM and in the higher-spin quantum sun models, their finite-size transition points remain quantitatively close when the local Hilbert-space dimension is kept fixed. This indicates that the location of the EBT is primarily controlled by the local Hilbert-space dimension. The higher spin results are therefore consistent with the same $q$-dependent hybridization prediction that describes the QPSM.

\end{document}


\title{\textcolor{titleblue}{Supplemental Material:}
\textcolor{titleblue}{Ergodicity breaking transition beyond qubits}
}
\author{Tanay Pathak$^{*,\dagger}$\,\,\href{https://orcid.org/0000-0003-0419-2583}
{\includegraphics[scale=0.05]{orcidid.pdf}}}
\affiliation{Department of Physics, Kyoto University, Kitashirakawa Oiwakecho, Sakyo-ku, Kyoto 606-8502, Japan}

\author{Abhik Kumar Saha$^{*,\ddagger}$\,\,\href{https://orcid.org/0000-0001-8168-6742}
{\includegraphics[scale=0.05]{orcidid.pdf}}}
\affiliation{Department of Physics, Kyoto University, Kitashirakawa Oiwakecho, Sakyo-ku, Kyoto 606-8502, Japan}

\begingroup
\renewcommand{\thefootnote}{*}
\footnotetext{These authors contributed equally to this work.}
\renewcommand{\thefootnote}{$\dagger$}
\footnotetext{Contact author: \href{mailto:pathak.tanay.4s@kyoto-u.ac.jp} {pathak.tanay.4s@kyoto-u.ac.jp}}
\renewcommand{\thefootnote}{$\ddagger$}
\footnotetext{Contact author: \href{mailto:saha.abhikkumar.3k@kyoto-u.ac.jp} {saha.abhikkumar.3k@kyoto-u.ac.jp}}
\endgroup
\maketitle
\onecolumngrid
\setcounter{figure}{0}
\renewcommand{\thefigure}{S\arabic{figure}}
\setcounter{figure}{0}
\renewcommand{\theequation}{S\arabic{equation}}
\setcounter{equation}{0}

In this supplemental material, we provide details on numerics and additional supporting results. In particular
\begin{itemize}
    \item In Section \ref{ssec:Htime}, we provide the method used to extract the Heisenberg time ($t_{\rm H}$).
    \item In Section \ref{ssec:thtime}, we provide the method used to extract the Thouless time ($t_{\rm th}$). 
    \item In Section \ref{ssec:datacol}, we provide the results on data collapse.
\end{itemize}

\section{Extraction of Heisenberg time ($t_{\rm H}$)}\label{ssec:Htime}
\begin{figure}[h]
    \centering
    \includegraphics[width=\columnwidth]{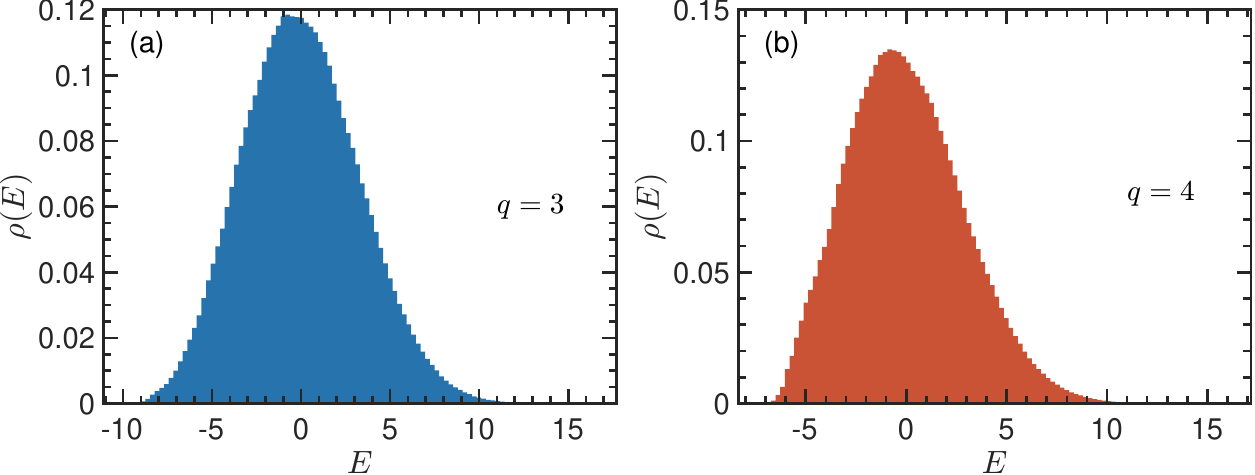}
    \caption{Many-body density of states $\rho(E)$ for the QPSM at $\alpha=0.70$: (a) $q=3$ with $N=2$ and $L=8$, and (b) $q=4$ with $N=2$ and $L=6$. In both cases, the density of states exhibits an approximately Gaussian profile, supporting the Gaussian approximation used in estimating the Heisenberg time $t_{\rm H}$ form the spectral variance $\Gamma_{0}$.}
    \label{sfig:dos_q_3_4}
\end{figure}

The Heisenberg time $t_{\rm H}$ is determined by the inverse of the characteristic many-body level spacing. Following Ref.~\cite{PhysRevLett.129.060602}, we estimate $t_{\rm H}$ from the width of the many-body density of states, characterized by the spectral variance $\Gamma_{0}$. We define $\Gamma_{0}$ as 
\begin{equation}
\Gamma_0^2 = \left\langle {\rm Tr}(H^2)/\mathcal{D} - {\rm Tr}(H)^2/\mathcal{D}^2\right\rangle_{d},
\end{equation}
where $\mathcal{D}$ is the total Hilbert-space dimension and $\langle\cdots\rangle_{d}$ denotes the disorder average. This quantity measures the typical spread of the eigenvalues around the mean energy. The characteristic level spacing is then estimated as $\Delta E=\Gamma_0/\chi \mathcal{D}$, where $\chi$ is the fraction of states lying in the interval $[\bar{E},\bar{E}+\Gamma_{0}]$, with $\bar{E}=\langle \mathrm{Tr}\{H\}\rangle/\mathcal{D}$. Fig.~\ref{sfig:dos_q_3_4} shows the many-body density of states of the quantum Potts sun model (QPSM) for $q=3$ and $q=4$ cases. In both cases, $\rho(E)$ exhibits an approximately Gaussian profile, with the spectral weight concentrated around the center of the spectrum and suppressed toward the edges. This supports the Gaussian approximation used in estimating the characteristic many-body level spacing from the spectral variance $\Gamma_{0}$. Assuming a Gaussian density of states \cite{FRENCH1970449,BOHIGAS1971383,Hartmann_2004,Hartmann_2005}, we obtain $\chi=(1/2)\rm{erf}(1/\sqrt{2})$, thus yielding Heisenberg time as
\begin{equation}
t_{\mathrm H}=\chi \mathcal{D}/\Gamma_0.
\end{equation}
Unlike the full spectral bandwidth $E_{\rm max}-E_{\rm min}$, which is determined by the spectral edges, $\Gamma_0$ is obtained from the first and second moments of the full spectrum and characterizes the typical spread of the eigenvalues about the mean energy. As a consistency check, we also estimated $t_{\rm H}$ directly from the inverse mean adjacent level spacing. When the spectral average is taken over the entire spectrum, the result is effectively controlled by the total bandwidth, since the sum of all adjacent spacings reduces to $E_{\rm max}-E_{\rm min}$. In contrast, restricting the calculation to the central part of the spectrum gives a value close to the $\Gamma_0$- based estimate. In our calculations, the two estimates differ by only about $5-6\,\%$ quantitatively, showing that the spectral-width prescription captures the same characteristic time scale as the direct bulk level-spacing estimate. We therefore use the $\Gamma_0$ based value of $t_{\rm H}$ to rescale the time axis in the spectral form factor (SFF) analysis.

\section{Extraction of Thouless time ($t_{\rm Th}$)}\label{ssec:thtime}
The Thouless time is extracted from the onset of universal random-matrix behavior in the SFF. For each disorder realization, we first unfold the many-body energy spectrum $\{E_n\}$, to obtain the unfolded eigenvalues $\{\epsilon_n\}$, for which the local mean level spacing is unity. Following Ref.~\cite{PhysRevE.102.062144}, the unfolding is performed separately for each realization using the cumulative spectral function
\begin{equation}
G(E)=\sum_n \Theta(E-E_n),
\end{equation}
which is fitted by a smooth polynomial. The unfolded eigenvalues are then obtained by evaluating the smooth part of the cumulative spectral function at the original eigenenergies. After unfolding, the scaled Heisenberg time is therefore $\tau_{\rm H}=1$. The SFF is calculated for each realization as described in the main text and subsequently averaged over disorder realizations.

The scaled Thouless time $\tau_{\rm Th}$ is identified from the onset of the universal GOE regime, where the numerical SFF approaches the GOE prediction. We use 
\begin{equation}
K_{\rm GOE}(\tau)=
\begin{cases}
2\tau-\tau\ln(1+2\tau), & \tau\leq1,\\[1mm]
2-\tau\ln\left(\dfrac{2\tau+1}{2\tau-1}\right), & \tau>1 .
\end{cases}
\end{equation}
Here $\tau=t/t_{\rm H}$ denotes the scaled time. For the numerical extraction, each $K(\tau)$ curve is evaluated at $5000$ values of $\tau$ in the interval $\tau\in[1/(2\pi\mathcal{D}),5]$, and the $\tau$ are equally spaced on a logarithmic scale. To suppress residual fluctuations, the disorder-averaged SFF is smoothed using a running mean over $150$ neighboring values of $K(\tau)$.

We quantify the deviation of the smoothed numerical SFF from the GOE prediction by
\begin{equation}
\Delta K(\tau)=\left|\log_{10}\left[\frac{K(\tau)}{K_{\rm GOE}(\tau)}\right]\right|,
\end{equation}
At short scaled times, $K(\tau)$ deviates from the GOE prediction, whereas upon entering the universal regime it approaches $K_{\rm GOE}(\tau)$. The scaled Thouless time $\tau_{\rm Th}$ is determined using   
\begin{equation}
\Delta K(\tau_{\rm Th})=\varepsilon ,
\end{equation}
Following Ref.~\cite{PhysRevLett.129.060602}, we determine $\tau_{\rm Th}$ by starting from the large $\tau$ regime and moving toward smaller $\tau$. Since residual fluctuations can lead to several crossings of the threshold, $\tau_{\rm Th}$ is taken as the last point for which $\Delta K(\tau)>\varepsilon$ when moving in this direction. This procedure identifies the onset of the regime in which the SFF remains within the prescribed tolerance of the GOE prediction and avoids selecting an accidental earlier crossing. Since $\tau_{\rm H}=1$, the ratio of the Heisenberg and Thouless times is equivalently
\begin{equation}
\frac{t_{\rm H}}{t_{\rm Th}}
=\frac{\tau_{\rm H}}{\tau_{\rm Th}}
=\frac{1}{\tau_{\rm Th}}.
\end{equation}
This ratio is used in the main text to characterize the finite-size behavior across the ergodicity-breaking transition. 
\begin{figure}[htbp]
    \centering
    \includegraphics[width=\columnwidth]{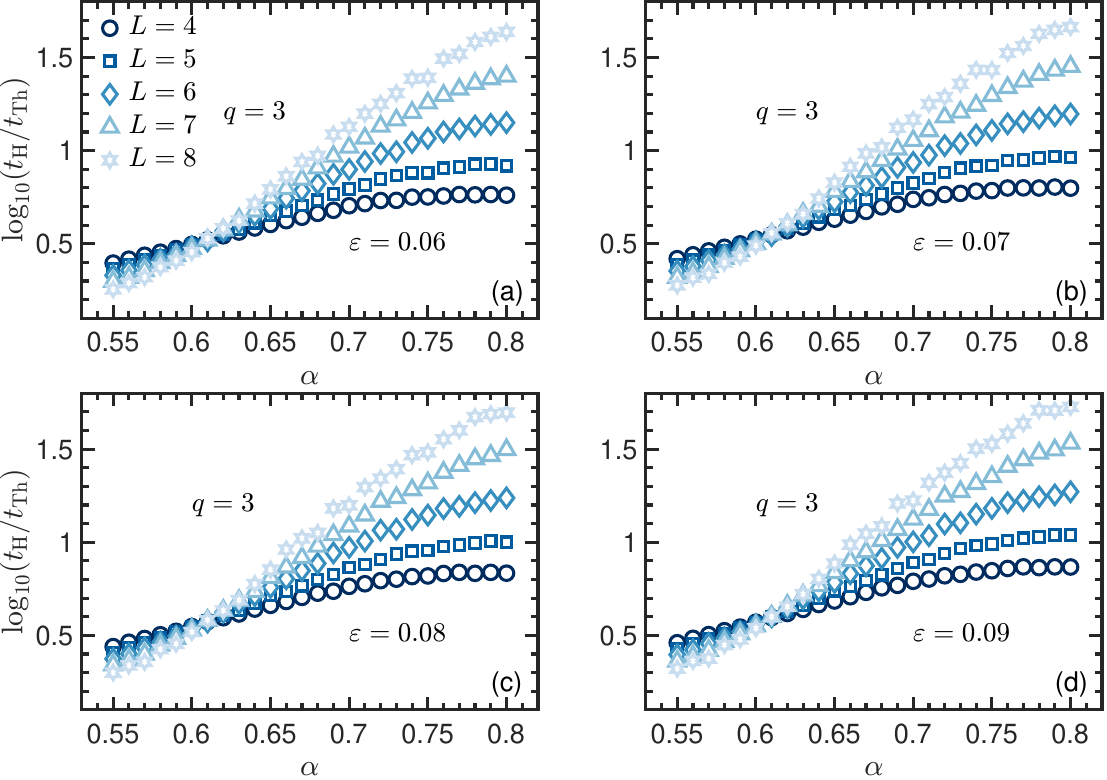}
    \caption{Dependence of $\log_{10}(t_{\rm H}/t_{\rm Th})$ on the threshold $\varepsilon$ for the QPSM model with $q=3$. Results are shown as a function of $\alpha$ for $L=4,5,6,7$ and $8$, with (a) $\varepsilon=0.06$, (b) $\varepsilon=0.07$, (c) $\varepsilon=0.08$, and (d) $\varepsilon=0.09$. The crossing of the different system sizes remains nearly unchanged for $\varepsilon=0.06-0.09$, demonstrating that the estimated transition point is robust against the choice of threshold $\varepsilon$.}
    \label{sfig:diff_epsilon_EBP_q_3}
\end{figure}

To examine the sensitivity of the extracted Thouless time $\tau_{\rm Th}$ to the choice of threshold $\varepsilon$, Fig.~\ref{sfig:diff_epsilon_EBP_q_3} shows $\log_{10}(t_{\rm H}/t_{\rm Th})$ as a function of $\alpha$ for $q=3$ in QPSM with different system sizes, using $\varepsilon=0.06,0.07,0.08$ and $0.09$. In all cases, the curves exhibit the same finite-size behavior: $\log_{10}(t_{\rm H}/t_{\rm Th})$ decreases with $L$ below the transition and increases with $L$ above it, with an approximately size-independent crossing near $\alpha\approx0.61$. The location of this crossing remains nearly unchanged over the range of $\varepsilon$ considered, showing that the estimated transition point is robust against the threshold used to determine $\tau_{\rm Th}$.

\begin{figure}[htbp]
    \centering
    \includegraphics[width=\columnwidth]{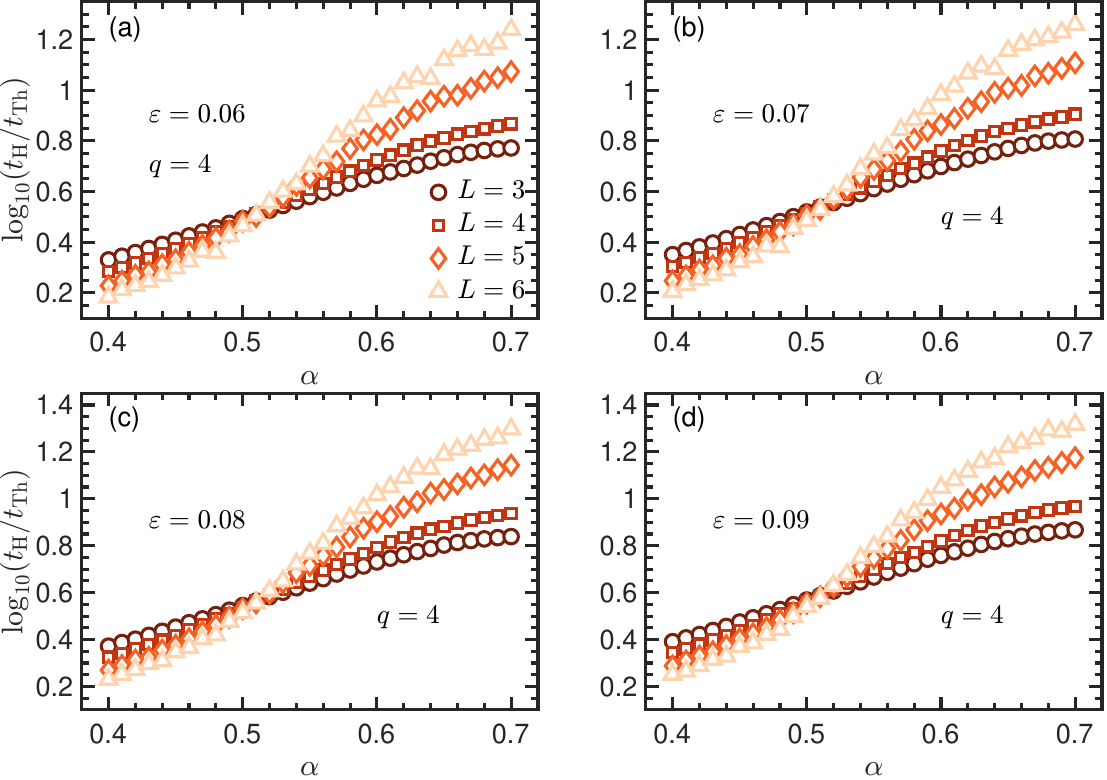}
    \caption{Dependence of $\log_{10}(t_{\rm H}/t_{\rm Th})$ on the threshold $\varepsilon$ for the QPSM model with $q=4$. Results are shown as a function of $\alpha$ for $L=3,4,5$ and $6$, with (a) $\varepsilon=0.06$, (b) $\varepsilon=0.07$, (c) $\varepsilon=0.08$, and (d) $\varepsilon=0.09$. The crossing of the different system sizes remains nearly unchanged for $\varepsilon=0.06-0.09$, demonstrating that the estimated transition point is robust against the choice of threshold.}
    \label{sfig:diff_epsilon_EBP_q_4}
\end{figure}

The same robustness is observed for $q=4$, as shown in Fig.~\ref{sfig:diff_epsilon_EBP_q_4}, where $\log_{10}(t_{\rm H}/t_{\rm Th})$ is plotted as a function of $\alpha$ for $L=3,4,5,6$ with different values of $\varepsilon$. For all threshold values, the curves exhibit the same finite-size trend and cross near $\alpha\approx0.52$. The crossing remains nearly unchanged over different values of $\varepsilon$, confirming that the estimated transition point is robust against the choice of threshold used to extract $\tau_{\rm Th}$.

\begin{figure}[htbp]
    \centering
    \includegraphics[width=\columnwidth]{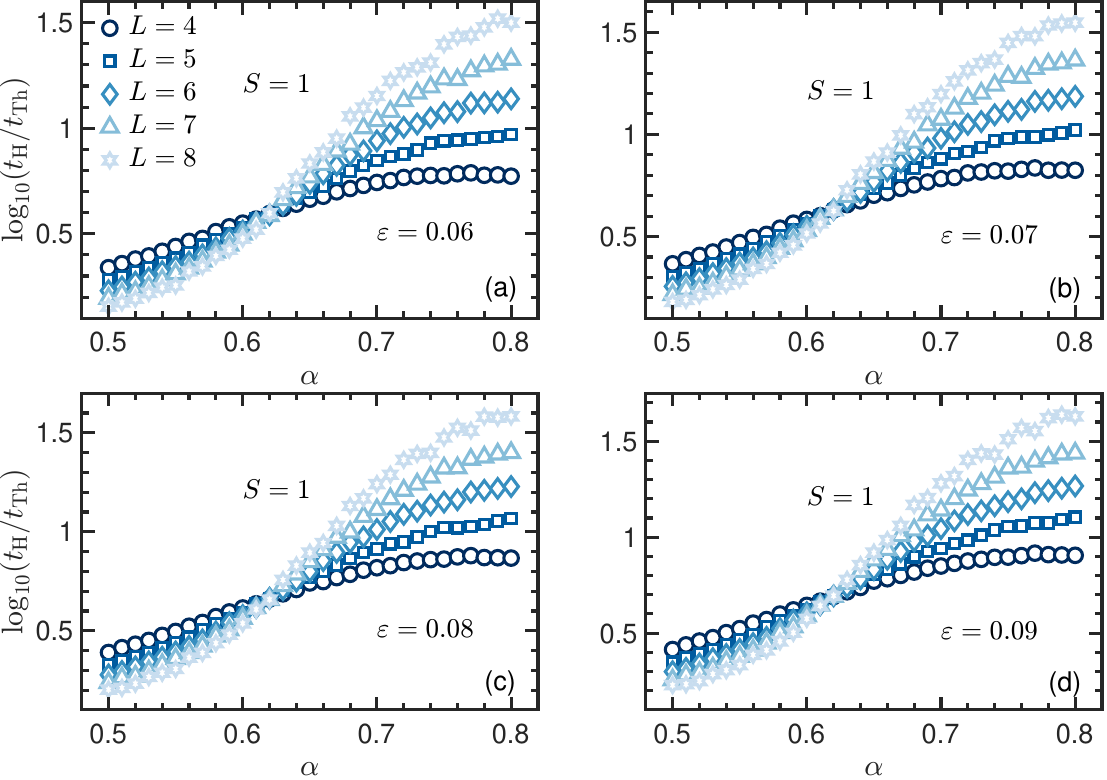}
    \caption{Dependence of $\log_{10}(t_{\rm H}/t_{\rm Th})$ on the threshold $\varepsilon$ for the spin-$1$ quantum sun model. Results are shown as a function of $\alpha$ for $L=4,5,6,7$ and $8$, with (a) $\varepsilon=0.06$, (b) $\varepsilon=0.07$, (c) $\varepsilon=0.08$, and (d) $\varepsilon=0.09$. The crossing of the different system sizes remains nearly unchanged for $\varepsilon=0.06-0.09$, demonstrating that the estimated transition point is robust against the choice of threshold.}
    \label{sfig:diff_epsilon_EBP_S_1}
\end{figure}

We next apply the same analysis to the spin-$1$ quantum sun model, for which the local Hilbert-space dimension is also three. Fig.~\ref{sfig:diff_epsilon_EBP_S_1} shows $\log_{10}(t_{\rm H}/t_{\rm Th})$ for $L=4,5,6,7,8$ with different values of $\varepsilon$. For all threshold values, the curves exhibit an approximately size-independent crossing near $\alpha\approx 0.62$, consistent with the transition point obtained for the $q=3$ QPSM. The persistence of the approximately same transition point despite the different local interaction structure supports the robustness of the ergodicity-breaking transition against microscopic details of the interaction present in the model at fixed local Hilbert-space dimension. 

\begin{figure}[htbp]
    \centering
    \includegraphics[width=\columnwidth]{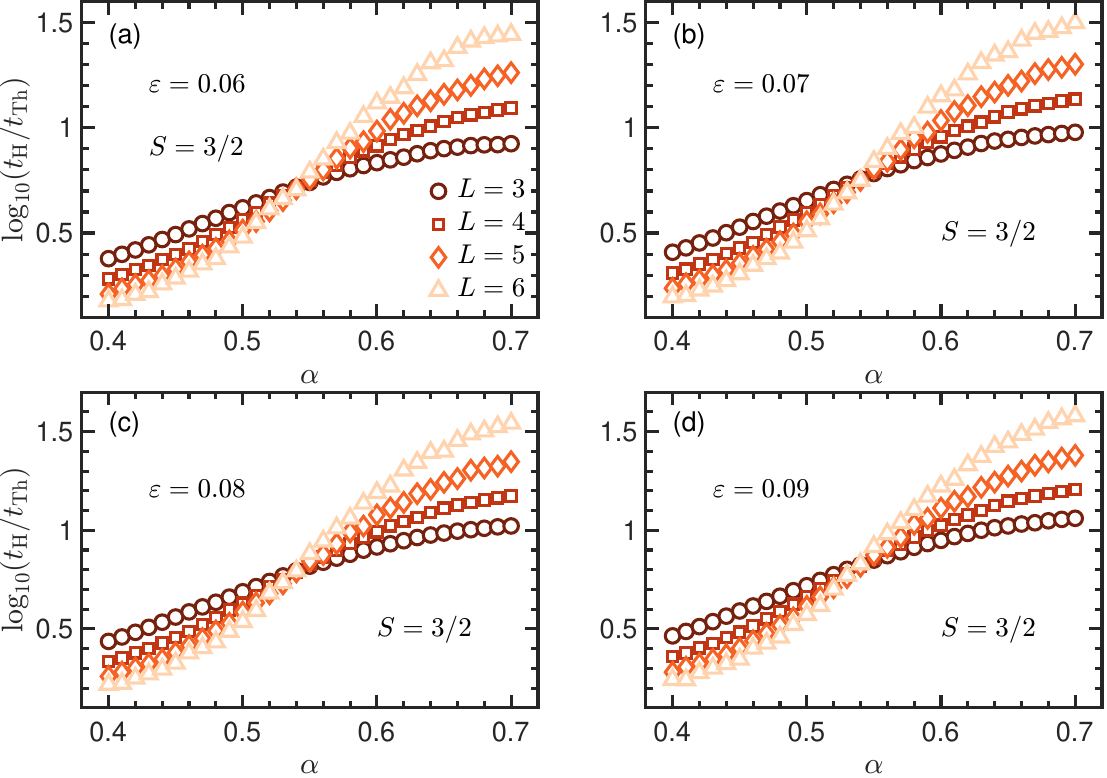}
    \caption{Dependence of $\log_{10}(t_{\rm H}/t_{\rm Th})$ on the threshold $\varepsilon$ for the spin-$3/2$ quantum sun model. Results are shown as a function of $\alpha$ for $L=3,4,5$ and $6$, with (a) $\varepsilon=0.06$, (b) $\varepsilon=0.07$, (c) $\varepsilon=0.08$, and (d) $\varepsilon=0.09$. The crossing of the different system sizes remains nearly unchanged for $\varepsilon=0.06-0.09$, demonstrating that the estimated transition point is robust against the choice of threshold. }
    \label{sfig:diff_epsilon_EBP_S_3by2}
\end{figure}

We finally repeat the same analysis for spin-$3/2$ quantum sun model, whose local Hilbert space dimension is four and therefore matches that of the $q=4$ QPSM. Fig.~\ref{sfig:diff_epsilon_EBP_S_3by2} shows $\log_{10}(t_{\rm H}/t_{\rm Th})$ as a function of $\alpha$ for different system sizes with different values of $\varepsilon$. For all threshold values, the curves exhibit an approximately size-independent crossing near $\alpha\approx0.54$, consistent with the transition point obtained for $q=4$ QPSM. Together with spin-$1$ results, this shows that the transition point remains nearly unchanged when the microscopic interaction structure is altered while the local Hilbert-space dimension is kept fixed. 

In the main text, the threshold $\varepsilon$ is kept fixed for a given local Hilbert-space dimension throughout the analysis. We use $\varepsilon=0.06$ for $q=3$ and $\varepsilon=0.08$ for $q=4$ for all system sizes and values of $\alpha$. These thresholds are chosen to remain above the residual
fluctuation level of $\Delta K(\tau)$ while reliably identifying the approach to the GOE regime. Once $\tau_{\rm Th}$ is determined, the Thouless time in physical units is defined as
\begin{equation}
t_{\rm Th}=\tau_{\rm Th}t_{\rm H}.
\label{seq:thouless_p}
\end{equation}
We therefore use the Eq.~\eqref{seq:thouless_p} to calculate the Thouless time $t_{\rm Th}$ in physical unit for further analysis.

\begin{figure}[htbp]
    \centering
    \includegraphics[width=\columnwidth]{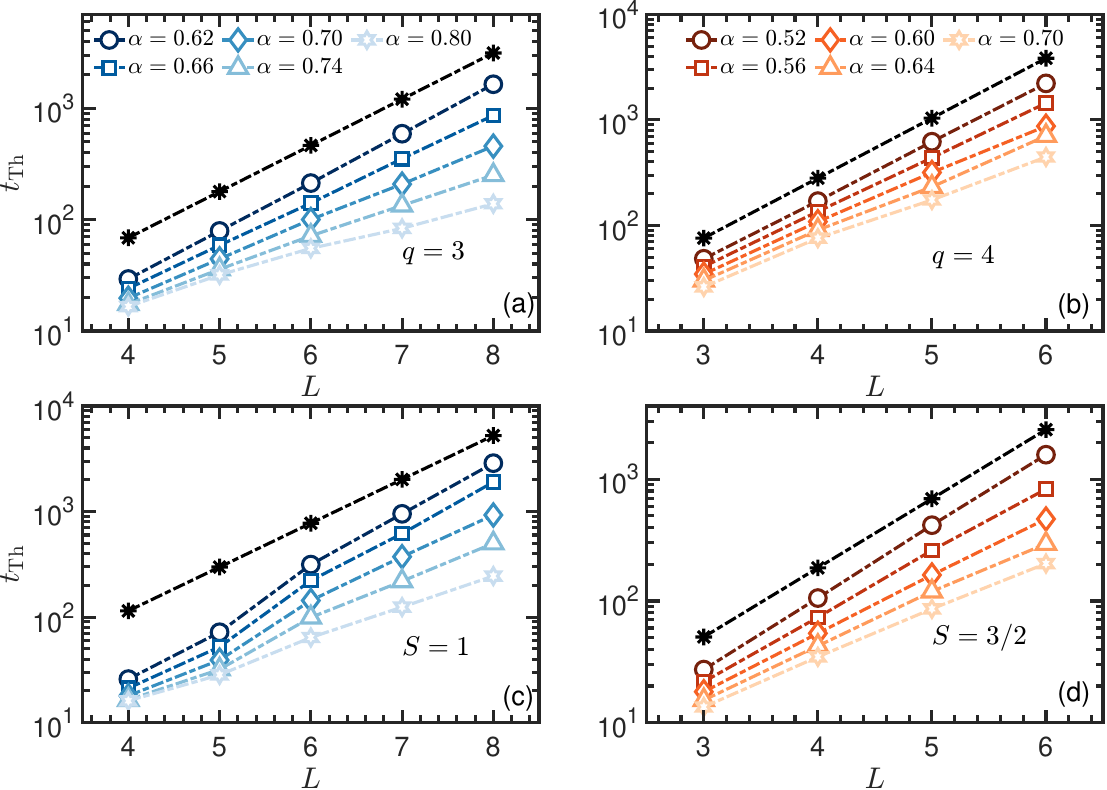}
    \caption{System-size dependence of the Thouless time $t_{\rm Th}$ for (a) the QPSM with $q=3$, (b) the QPSM with $q=4$, (c) the spin-$1$ quantum sun model, and (d) the spin-$3/2$ quantum sun model. The symbols show the numerical $t_{\rm Th}$ for the different values of $\alpha$ indicated in (a) and (b). The black dashed-dotted lines with markers show the analytical estimate $t_{\rm Th}^{*}=\exp[\ln(1/\alpha^2)L]$. For better visibility, the analytical curves are multiplied by factors $1.5$, $1.5$, $2.5$, and $1.5$ in panels (a)-(d), respectively. For this reference curve, we use the smallest value of $\alpha$ considered in each set, namely $\alpha=0.62$ for $q=3$ and $S=1$, $\alpha=0.52$ for $q=4$ and $S=3/2$. The approximately linear behavior on the semi-logarithmic scale shows the exponential growth of $t_{\rm Th}$ with $L$, with a growth rate that decreases as $\alpha$ increases. Note that the $y$-axes are in logarithmic scale.}
    \label{sfig:data_thouless_time}
\end{figure}

Fig.~\ref{sfig:data_thouless_time} shows the resulting system-size dependence of $t_{\rm Th}$ for the QPSM with $q=3$ and $q=4$, together with the corresponding spin-$1$ and spin-$3/2$ quantum sun models. In all four cases, $t_{\rm Th}$ increases exponentially with $L$, as reflected by the nearly linear dependence on the semi-logarithmic scale. The growth is higher for the smallest $\alpha$ and becomes progressively weaker as $\alpha$ increases. For comparison, the black dashed curves with markers show the analytical estimate \cite{PhysRevLett.129.060602}
\begin{equation}
t_{\rm Th}^{*}=\exp[\ln(1/\alpha^2)L],   
\end{equation}
evaluated at the smallest $\alpha$ considered in each set: $\alpha=0.62$ for $q=3$ and $S=1$, and $\alpha=0.52$ for $q=4$ and $S=3/2$. The numerical data exhibit the same exponential finite-size trend as the analytical estimate, while the reduction of the slope with increasing $\alpha$ shows that the Thouless time grows more slowly as the system moves further into the ergodic regime. 

\section{Critical exponents and data collapse}\label{ssec:datacol}
Close to the ergodicity-breaking transition, a characteristic length scale $\xi$ is expected to diverge as
\begin{equation}
\xi \sim |\alpha-\alpha_{c}|^{-\nu},    
\end{equation}
where $\nu$ is the corresponding critical exponent and $\alpha_{c}$ is the transition point. In a finite-size system, the divergence of $\xi$ is limited by the system size $L$, and the crossover region is therefore determined by the competition between these two length scales. Within the standard finite-size scaling framework \cite{PhysRevB.91.081103,PhysRevX.7.021013,PhysRevB.97.220201,PhysRevB.97.201105}, this leads to the rescaled variable 
\begin{equation}
x=(\alpha-\alpha_c)L^{1/\nu}.
\label{seq:scaling_variable}
\end{equation}
An observable $\mathcal{O}(L,\alpha)$ probing the transition is then assumed to satisfy the scaling form 
\begin{equation}\mathcal{O}(L,\alpha)=
F_{\mathcal{O}}\!\left[(\alpha-\alpha_c)L^{1/\nu}\right],
\label{seq:collapse}
\end{equation}
where $F_{\mathcal{O}}$ is the corresponding scaling function. We apply this scaling analysis independently to the disorder-averaged gap ratio $\langle r\rangle_{d}$ and to $\log_{10}(t_{\rm H}/t_{\rm Th})$ for a finite-size estimate of the transition point. If the finite-size scaling hypothesis is valid, data obtained for different system sizes $L$ should collapse onto a common curve when plotted as a function of $x$ within the critical region. In particular, $\alpha=\alpha_{c}$ corresponds to $x=0$ for every system size. We extract $\alpha_c$ and $\nu$ by optimizing the quality of this collapse. For each trial pair $(\alpha_{c},\nu)$, we first restrict the data to a finite interval of $\alpha$ centered around the crossing region and calculate
\begin{equation}
x_{L,k}=(\alpha_{L,k}-\alpha_c)L^{1/\nu},
\end{equation}
where $k$ labels the data points for a given $L$. Since the values of $x_{L,k}$ generated for different system sizes do not generally coincide, the comparison between the curves is restricted to their common scaling interval,
\begin{equation}
x_{\min}=\max_L\!\left[\min_k(x_{L,k})\right],\qquad
x_{\max}=\min_L\!\left[\max_k(x_{L,k})\right].
\end{equation}
Within this interval, each curve is linearly interpolated onto a common grid of $N_x=500$ equally spaced values of $x_{j}$. For a generic observable $\mathcal{O}$, we define the mean interpolated curves as 
\begin{equation}
\overline{\mathcal{O}}(x_j)=\frac{1}{N_L}\sum_L \mathcal{O}_L(x_j),
\end{equation}
where $N_{L}$ denotes the number of system sizes included in the analysis. We then quantify the quality of the collapse by the mean-square spread of the individual curves around $\overline{\mathcal{O}}$,
\begin{equation}
\mathcal{C}(\alpha_c,\nu)=\frac{1}{N_LN_x}\sum_L\sum_{j=1}^{N_x}\left[\mathcal{O}_L(x_j)-\overline{\mathcal{O}}(x_j)\right]^2 .
\label{seq:collapse_cost}
\end{equation}

Thus, $\mathcal{C}$ measures the residual finite-size dependence after rescaling: a smaller value corresponds to a better overlap of the curves and hence to a better realization of the scaling form in Eq.~\eqref{seq:collapse}. The optimal values of $\alpha_{c}$ and $\nu$ are determined by minimizing $\mathcal{C}(\alpha_c,\nu)$ using a differential-evolution algorithm, followed by a local refinement of the minimum. As a consistency check, we repeated the analysis using the alternative cost function of Refs.~\cite{PhysRevLett.129.060602,PhysRevB.102.064207,PhysRevB.109.L180201} and obtained nearly identical values of $\alpha_{c}$ and $\nu$, confirming the robustness of the finite-size estimates. The optimization is restricted to a narrow interval of $\alpha$ around the transition, where the scaling form is expected to apply, following the same strategy commonly used in finite-size analyses near ergodicity-breaking transitions \cite{PhysRevE.102.062144,PhysRevB.91.081103,PhysRevLett.129.060602}. The procedure is carried out separately for $\langle r\rangle_{d}$ and $\log_{10}(t_{\rm H}/t_{\rm Th})$, providing independent finite-size estimates of the transition point and the exponent $\nu$. Using the optimized values of $\alpha_{c}$ and $\nu$, the data for different system sizes collapse onto a common curve when plotted as a function of $(\alpha-\alpha_c)L^{1/\nu}$. For all the models, the collapse is particularly good in the critical region $-0.5\lesssim x\lesssim0.5$, supporting the finite-size scaling description near the ergodicity-breaking transition.

\begin{figure}[htbp]
    \centering
    \includegraphics[width=\columnwidth]{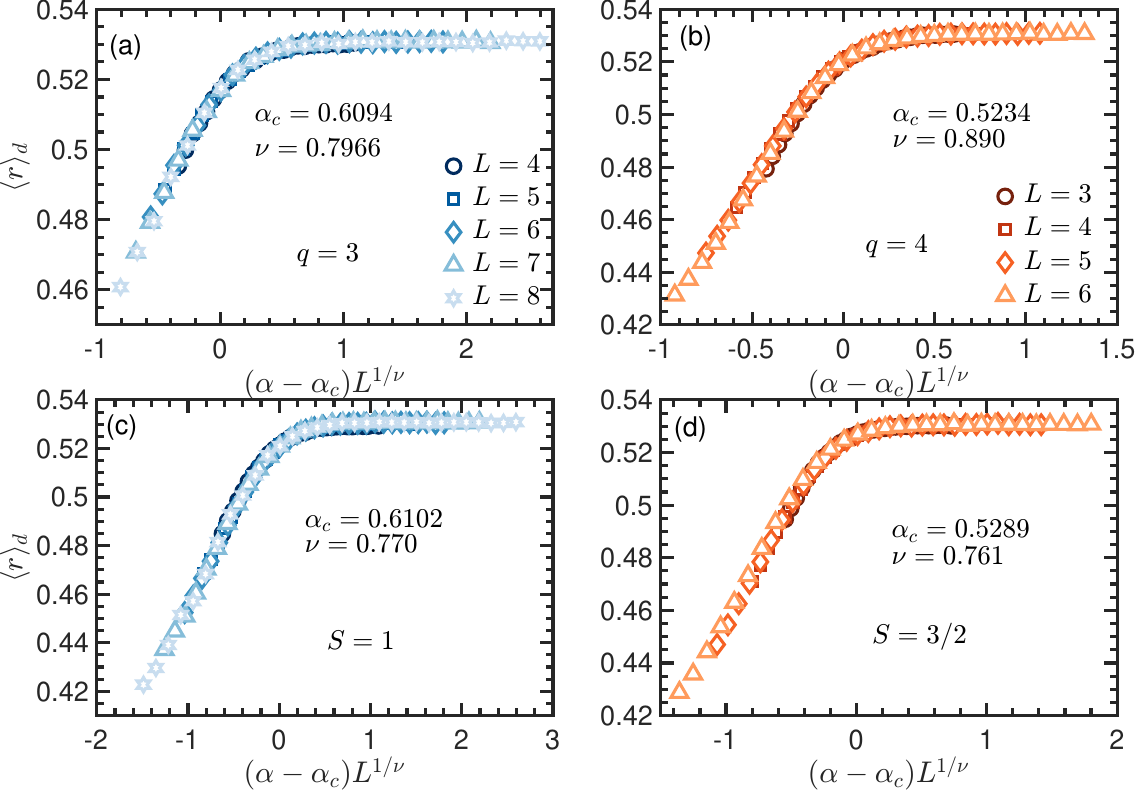}
    \caption{Data collapse of the disorder-averaged gap ratio $\langle r \rangle_{d}$ as a function of the rescaled variable $(\alpha-\alpha_{c})L^{1/\nu}$. The critical coupling $\alpha_{c}$ and exponent $\nu$ are obtained by optimizing the collapse of the different system sizes. (a) QPSM with $q=3$, $L=4,5,6,7,8$ yielding $\alpha_{c}=0.6094$ and $\nu=0.7966$. (b) QPSM with $q=4$, $L=3,4,5,6$ yielding $\alpha_{c}=0.5234$ and $\nu=0.890$ (c) Spin-$1$ quantum sun model with the same system sizes as (a) yielding $\alpha_{c}=0.6102$ and $\nu=0.770$. (d) Spin-$3/2$ quantum sun model with the same system sizes as (b) yielding $\alpha_{c}=0.5289$ and $\nu=0.761$. Different symbols denote different system sizes $L$. The collapse of the different system sizes onto a common scaling curve in the vicinity of the transition provides finite-size estimates of $\alpha_{c}$ and $\nu$.  }
    \label{sfig:data_collapse_ratio}
\end{figure}

Fig.~\ref{sfig:data_collapse_ratio} shows the finite-size scaling collapse of the disorder-averaged gap ratio $\langle r \rangle_{d}$ for the $q=3$ and $q=4$ QPSM and for the corresponding spin-$1$ and spin-$3/2$ quantum sun models. In all four cases, the data for different system sizes collapse well onto a common curve in the vicinity of the transition. The optimized transition points are $\alpha_c=0.6094$ and $0.5234$ for $q=3$ and $q=4$ QPSM, respectively, while the corresponding higher-spin quantum sun models give $\alpha_c=0.6102$ for $S=1$ and $\alpha_c=0.5289$ for $S=3/2$. Thus, models with the same local Hilbert-space dimension yield very similar transition points despite their different microscopic interaction structures. The scaling collapse is consistent with the crossing observed in the unscaled gap-ratio data and allows us to extract finite-size estimates of $\alpha_{c}$ and $\nu$. 

\begin{figure}[htbp]
    \centering
    \includegraphics[width=\columnwidth]{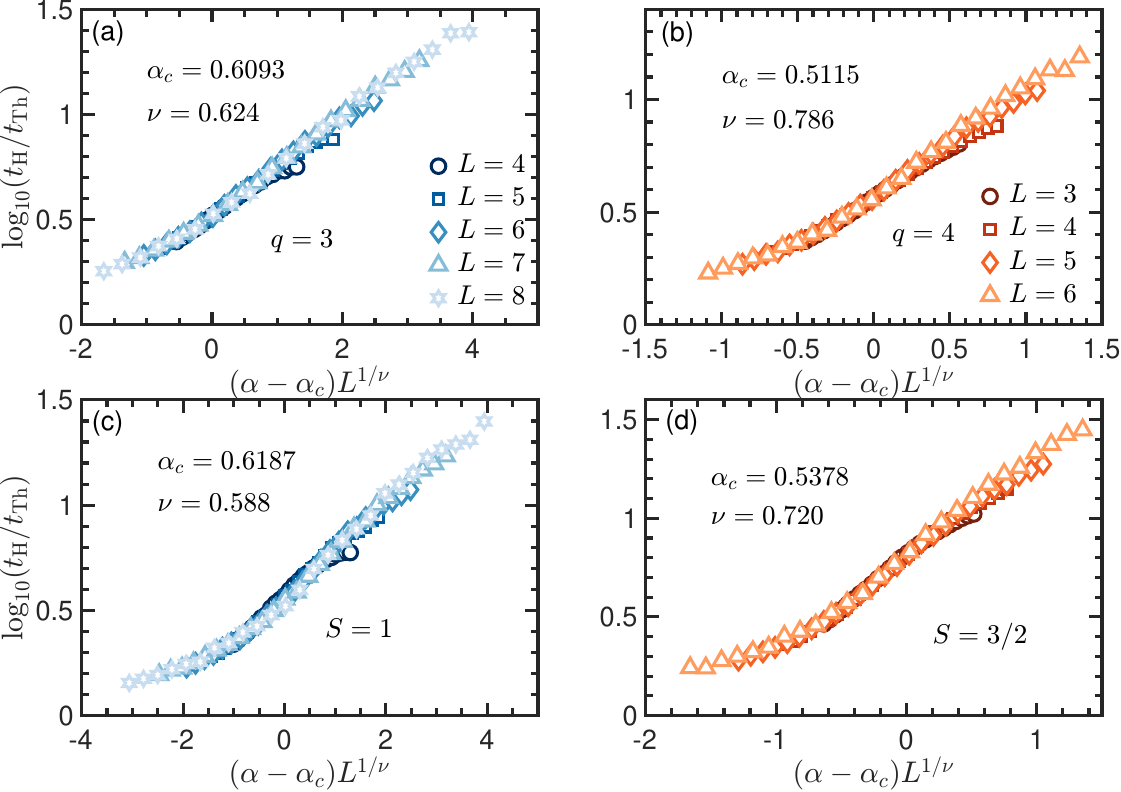}
    \caption{Data collapse of the $\log_{10}(t_{\rm H}/t_{\rm Th})$ as a function of the rescaled variable $(\alpha-\alpha_{c})L^{1/\nu}$. The critical coupling $\alpha_{c}$ and exponent $\nu$ are obtained by optimizing the collapse of the different system sizes. (a) QPSM with $q=3$, $L=4,5,6,7,8$ yielding $\alpha_{c}=0.6093$ and $\nu=0.624$. (b) QPSM with $q=4$, $L=3,4,5,6$ yielding $\alpha_{c}=0.5115$ and $\nu=0.786$ (c) Spin-$1$ quantum sun model with same system sizes as (a) yielding $\alpha_{c}=0.6187$ and $\nu=0.588$. (d) Spin-$3/2$ quantum sun model with same system sizes as (b) yielding $\alpha_{c}=0.5378$ and $\nu=0.720$. Different symbols denote different system sizes $L$. The collapse of the different system sizes onto a common scaling curve in the vicinity of the transition provides finite-size estimates of $\alpha_{c}$ and $\nu$. }
    \label{sfig:data_collapse_EBP}
\end{figure}

A complementary scaling analysis of $\log_{10}(t_{\rm H}/t_{\rm Th})$ is shown in Fig.~\ref{sfig:data_collapse_EBP}. Again, the data for different system sizes collapse onto a common curve near the transition for all four models. The resulting transition points are $\alpha_c=0.6093$ for $q=3$, $\alpha_c=0.5115$ for $q=4$, $\alpha_c=0.6187$ for $S=1$, and $\alpha_c=0.5378$ for $S=3/2$, respectively. Importantly, the transition points obtained from this long-range spectral measure are close to those extracted independently from the gap-ratio data collapse. In particular, both analyses yield transition points near $\alpha_c \approx 0.61$ for $q=3$ QPSM and spin-$1$ quantum sun model, while for the $q=4$ QPSM and spin-$3/2$ quantum sun model the corresponding values lie around $\alpha_c\approx0.52-0.54$. The agreement between the two spectral measures, together with the close correspondence between the QPSM and higher-spin quantum sun models at fixed local Hilbert-space dimension, supports the robustness of the finite-size estimate of the ergodicity-breaking transition.

\bibliography{references}